\documentclass[aps,prl,twocolumn,nofootinbib,groupedaddress,amsfonts,floatfix]{revtex4-1}

\usepackage{graphicx,amsmath,amssymb,amstext}
\usepackage{amssymb,amsbsy,amsfonts,amsthm}
\usepackage{comment}

\usepackage{color}
\usepackage{ifthen}
\newboolean{editorial}
\setboolean{editorial}{true}
\newcommand{\editorial}[2]{\ifthenelse{\boolean{editorial}}{\textcolor{red}{[\textsf{\textbf{{#1}}}: }\textcolor{blue}{\textsf{{#2}}}\textcolor{red}{]}}{}}

\begin{document}

\title{Preheating with Nonminimal Kinetic Terms}

\author{Hillary L. Child${}^1$}
\author{John T. Giblin, Jr${}^{1,2}$}
\author{Raquel H. Ribeiro${}^2$}
\author{David Seery${}^3$}

\affiliation{${}^1$Department of Physics, Kenyon College, Gambier, OH 43022, USA}
\affiliation{${}^2$Department of Physics, Case Western Reserve University, Cleveland, OH 44106, USA}
\affiliation{${}^3$Astronomy Centre, University of Sussex, Falmer, Brighton, BN1 9QH, UK}

\begin{abstract}
We present the first (3 + 1)-dimensional numerical simulations of scalar fields with nonminimal kinetic terms.  As an example, we examine the existence and stability of preheating in the presence of a Dirac-Born-Infeld inflaton coupled to a canonical matter field.  The simulations represent the full nonlinear theory in the presence of an expanding universe.  We show that parametric resonance in the matter field along with self-resonance in the inflaton 
repopulate the universe with 
matter particles as efficiently as in traditional preheating.
\end{abstract}

\maketitle

Scalar fields need not be Klein-Gordon-like in four dimensions;  theories that include extra dimensions, whether large or small, have effective four-dimensional  kinetic terms that are nonminimal.  This has led to an increased interest in the dynamics of nonminimal models both to describe inflation, e.g. Dirac-Born-Infeld (DBI) inflation \cite{Silverstein:2003hf}, and even more recently, dark energy \cite{deRham:2011by,Hassan:2011zd,Nicolis:2008in}.  In this Letter, we will focus on DBI inflation, as an example of a system for which we have a scalar degree of freedom whose behavior is self-consistent and stable and whose nonminimal behavior is central to the dynamics of the model.  Here, we focus on what happens at the end of inflation when the scalar inflaton becomes inhomogeneous and its couplings to other fields are important.  To our knowledge, these are the first three-dimensional lattice simulations of this type of field theory.

Preheating \cite{Traschen:1990sw,Kofman:1994rk,GarciaBellido:1997wm,Khlebnikov:1997di,Greene:1997ge,Parry:1998pn,Bassett:1998wg,Easther:1999ws,Liddle:1999hq,Finelli:2001db,Bassett:2005xm,Podolsky:2005bw,GarciaBellido:1998gm} provides a mechanism by which the cold postinflationary universe can quickly and efficiently transfer energy into a matter sector, via a period of parametric resonance or a regime of tachyonic instability.  For the most part, studies of preheating have focused on the existence and stability of these processes in the presence of different inflationary potentials, e.g., Refs.\cite{Boyanovsky:1996ks,Tkachev:1996rv,Baacke:1996kj,Kofman:1997ga,GarciaBellido:1998gm,Mazumdar:2012md}, multiple fields \cite{Battefeld:2009xw}, and multiple decay channels \cite{Giblin:2010sp}.  DBI inflation might end with a coherently oscillating scalar field, 
and it has been unclear whether preheating can persist in the presence of nonlinear terms in its equation of motion.  The search for preheating in nonminimal models began in the work of Ref. \cite{Lachapelle:2008sy}, where the authors showed that a canonical scalar field can enhance the effects of parametric resonance in a coupled nonminimal matter field. 

The first studies that considered nonminimal inflationary fields \cite{Matsuda:2008hk} looked at the production of particles during the first oscillation, {\sl instant preheating}, as a possible source of non-Gaussianities in the cosmic microwave background (CMB); restricting the treatment to the first oscillations allows a complete analytic treatment of the amplification of the matter field but does not capture the majority of the preheating dynamics.  Later, the authors of Ref. \cite{Bouatta:2010bp} studied a DBI-type model with perturbative departures from a canonical kinetic term so that the potential dominated the preheating dynamics. In this case, preheating persists and is perturbatively enhanced, but in addition self-resonance occurs in the inflaton. At the other extreme, the authors of Ref. \cite{Karouby:2011xs} studied a case where the nonminimal kinetic term is dominant. In this case, the potential is irrelevant during most of the oscillation and the field profile develops a ``sawtooth'' oscillatory profile.  Furthermore, the authors argue that this feature suppresses preheating.

Here, we study the full nonlinear theory on a three-dimensional lattice.  We will not assume that the nonminimal terms are small or that the inflaton is purely homogenous.  Such a treatment allows us to study the existence of resonance in the coupled field, any self-resonance in the inflaton, and any nonlinear effects that take place after the resonant periods cease.  Furthermore, all analytical analyses of preheating with nonminimal kinetic terms require many simplifying assumptions: they may disregard Hubble friction and must ignore the spatial gradient terms in the equation of motion. Numerical simulations are needed to verify the possibility of preheating in this model outside the nonrelativistic limit and to see any interesting physics that arise from the interactions between the two fields.  We begin by considering the standard action for a DBI inflaton \cite{Silverstein:2003hf} 
\begin{equation}
\label{action}
S_\phi=\frac{1}{2} \int d^4x \sqrt{-g} \left[\frac{m_{\rm pl}^2}{8\pi}R+2P(\phi,X)\right],
\end{equation}
where $m_{\rm pl}^2=1/G$, $X= -\partial^\mu\phi\partial_\mu \phi$, and
\begin{equation}
P( \phi, X) = -\frac{1}{f(\phi)}\left(\sqrt{1-f(\phi)X} -1\right)- V(\phi).
\end{equation}
We will take the warp factor $f$ to satisfy \cite{Silverstein:2003hf}
\begin{equation}
f(\phi) = \frac{\lambda}{\left(\phi^2+\mu^2\right)^2}.
\end{equation}
It is related to the geometry of the extradimensional space in which the brane is moving; the form in Eq.~\ref{action} corresponds to the case of a cutoff anti-de Sitter throat. 
In a braneworld scenario, $\phi$ would correspond to the radial distance of the brane to the horizon. 
Note that in the limit where $fX \ll 1$, we recover the canonical kinetic term; the magnitude $|fX|$ therefore characterizes departures from the standard scenario.

The full nonlinear equation of motion for $\phi$, from Eq.~\eqref{action}, is
\begin{equation}
\begin{split}
\ddot{\phi}&\left(1+\frac{f}{a^2}\partial_j \phi \partial_j \phi \right) =
\\& - 3\frac{\dot{a}}{a}\frac{\dot{\phi}}{\gamma^2}+  \frac{\nabla^2 \phi}{a^2}+\frac{f^\prime}{f^2}-\frac{3f^\prime X}{2f} - \frac{1}{\gamma^3}\left(\frac{\partial V}{\partial \phi}  + \frac{f^\prime}{f^2} \right) 
\\&+ \frac{f}{a^2}\left[\left(\frac{\nabla^2\phi}{a^2}- \dot{\phi}\frac{\dot{a}}{a}\right)\partial_j\phi\partial_j\phi -\nabla^2\phi \dot{\phi}^2\right.  \\&\hphantom{{}=+ \frac{f}{a^2}} - \left. \frac{\partial_i\phi\partial_j \phi \partial_j \partial_i \phi}{a^2} + 2 \partial_j \phi \dot{\phi} \partial_j \dot{\phi}\right]
\end{split}
\label{DBIeom}
\end{equation}
in a Friedmann-Lema\^itre-Robertson-Walker universe, 
\begin{equation}
ds^2 = g_{\mu\nu}dx^\mu dx^\nu = -dt^2 + a^2(t)\left[dx^2+dy^2+dz^2\right],
\end{equation}
background.
The combination $\gamma = \left(1-fX\right)^{-1/2}$ describes how relativistically the brane is moving and is analogous to the usual Lorentz factor.  We will assume that the matter field is a massless canonical, Klein-Gordon field whose equation of motion is
\begin{equation}
\ddot{\chi} + 3\frac{\dot{a}}{a}\dot{\chi} - \frac{\nabla^2\chi}{a^2} + \frac{\partial V}{\partial \chi} = 0,
\end{equation}
and the full potential of the model is the combination of the inflationary potential and a coupling between the two fields,
\begin{equation}
V(\phi,\chi) = \frac{1}{2}m^2\phi^2 + \frac{1}{2} g^2 \phi^2 \chi^2.
\end{equation}
Since we are interested in effects due to the nonminimal nature of the field, we will not allow the parameters in the potential to change from the standard values used in the preheating literature, $m=10^{-6}\,{m_{\rm pl}}$ and $g^2 = 2.5\times10^{-5}$.  We will take the ratio $\lambda/\mu^4$ as a free parameter.

The contribution to the energy density from the inflaton, 
\begin{equation}
\label{rhophi}
\rho_\phi = \frac{1}{f}(\gamma^{-1}-1) + \gamma \dot{\phi}^2,
\end{equation} 
differs from the energy density associated with the matter field,
\begin{equation}
\rho_\chi = \frac{1}{2}\dot{\chi}^2 + \frac{\left(\nabla \chi \right)^2}{2a^2}.
\end{equation}
The combination of the two with the potential energy $\rho = \rho_\phi + \rho_\chi + V(\phi,\chi)$ together source the self-consistent expansion of the universe,
\begin{equation}
\frac{\dot{a}}{a} = \sqrt{\frac{8\pi}{3m_{\rm pl}^2}}\rho.
\end{equation}

{\sl Numerical simulations.--} 
Simulations were performed using Grid and Bubble Evolver ({\sc gabe}) \cite{GABE}, a new lattice evolution program that uses a second-order Runge-Kutta method of integration. Unlike existing lattice codes \cite{Felder:2000hq,Frolov:2008hy,Easther:2010qz,Huang:2011gf} that use symplectic integration routines, {\sc gabe} stores the field and field derivative value at the same times during the time step.  This method is slower and requires more physical memory to run than symplectic integrators but is necessary when terms include the product of the field with its time derivative.  We offset the slower performance by parallelization using {\sc openmp}.  We postpone a complete analysis of all relevant values of $\lambda/\mu^4$ for a subsequent publication, instead focusing on a single case, $\mu = 10^3\,m_{\rm pl}$ and $\lambda/\mu^4 = 5000 \: m^{-2} \: m_{\rm pl}^{-2} = 5 \times 10^{15} \: m_{pl}^{-4}$, as a representative model.  In the numerical simulations here, we use a box of $N^3=256^3$ points.  Running on four cores, each run takes approximately 8 days to complete.

We initialize the homogeneous mode of the matter field as $\chi(\vec{k}=0,t=0)=0$, whereas the homogeneous modes of the inflaton and its derivative are consistent with their values at the end of inflation, when $\ddot{a}$ drops below zero. We calculate these initial conditions using a sixth-order Runge-Kutta homogeneous field evolution program to evolve the field $\phi$ during inflation, when it obeys Eq.~\eqref{DBIeom} without the gradient terms.  For our case of interest here, $\phi_0 \approx 4.5\times 10^{-2}\,m_{\rm pl}$ and $\dot{\phi}_0 \approx -1.4\times10^{-8}\,m_{\rm pl}^2$.  Although the dynamics of the field is different, even at the end of inflation, these values are close to those expected in the canonical model.

Both fields and their derivatives are initialized with fluctuations that are consistent with the Bunch-Davies vacuum, in a way analogous to the situation in Ref. \cite{Felder:2000hq}, so that the power spectrum of the fields on the initial slice is
\begin{equation}
\left<|\phi(k)|^2\right> =  \frac{1}{2 \omega_{\phi,k}},\,\left<|\chi(k)|^2\right> =  \frac{1}{2 \omega_{\chi,k}}.
\end{equation}
We use this choice of vacuum for subhorizon fluctuations of the inflaton, as an approximation.  At the end of inflation, field gradients are washed out  and we expect fluctuations to be perturbations to the homogeneous background.  As there is no better choice, we assume a Bunch-Davies vacuum for both fields and expect that deviations from this choice of initial conditions will have no effect on the main result here, as preheating is generally insensitive to modifications to the initial power spectrum \cite{Frolov:2008hy}.
After initialization, the fields are evolved for several hundred oscillations of the inflaton field, until $t_f \sim 250\,m^{-1}$, in an expanding background. 

The software calculates the means, variances, and energy density of both fields as well as their energy and power spectra. Two-dimensional slices of the field profile are also generated as well as the space-averaged value of $\gamma$. The scale factor grows by a factor of about $a_f\sim17$ (when $\lambda/\mu^4=5000$) compared to $a_f\sim 27$ (in the canonical case for the same final time).

\begin{figure}[b]
\centering
\includegraphics[width=\columnwidth]{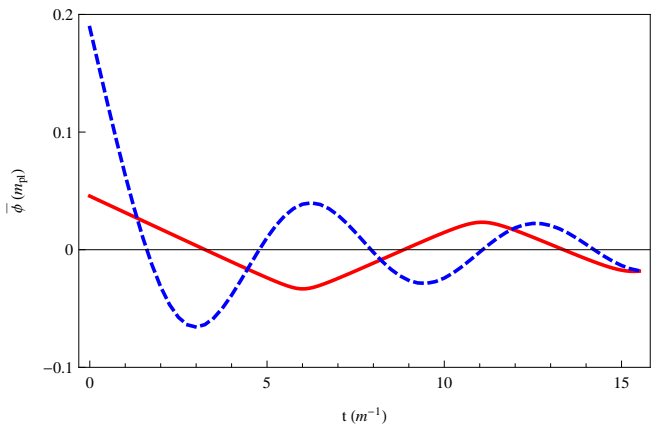}
\caption{Evolution of the mean inflaton field value for simulations with $\lambda/\mu^4=5000$, $\gamma_0\approx8.7$ (red solid line) and the canonical case $f\rightarrow 0$ (blue dashed line).}
\label{fig:means}
\end{figure}

{\sl Results.--} 
The evolution of the inflaton field during both inflation and its first few oscillations shows the effects expected due to the presence of these noncanonical terms. In the DBI case, as the velocity of the field grows, the relativistic factor $\gamma$ also grows, decreasing the influence of the potential term on the acceleration of the field.  Mathematically, to guarantee the reality of $\gamma$, $\dot{\phi}$ cannot exceed $f^{-1/2}$; however, in practice we see that the true speed limit is somewhat smaller and is a consequence of the nonlinear dynamics of the field.  We can easily see the effects of the speed limit on the oscillations of the inflaton after inflation. 
In fact, the mean value of $\phi$ on the lattice does not oscillate sinusoidally but instead reaches its maximum velocity quickly, after which it travels at a constant speed for much of each oscillation. The evolution of $\phi$ thus approaches a sawtooth pattern as $\lambda/\mu^4$ increases, as predicted in Ref. \cite{Karouby:2011xs}; see Fig. \ref{fig:means}. 

We have run simulations with initial values of $\gamma$ varying from $1 \lesssim \gamma \lesssim 10$, consistent with the Planck 2013 constraint $\gamma \lesssim 14$ at $95\%$ confidence \cite{Ade:2013uln}.
In all simulations, we see significant effects due to parametric resonance by approximately $t \sim 100\,m^{-1}$.  We can identify parametric resonance by the exponential amplification of particular modes of the matter field $\chi$, which result in an exponential increase in the variance of the matter field over time.  We can see this schematically by noting the inflaton is a coherently oscillating field $\phi = \Phi(t)$; the mode equations for the matter field,
\begin{equation}
\label{chimode}
\ddot{\chi}_{\vec{k}} + 3\frac{\dot{a}}{a}\chi_{\vec{k}} + \left(k^2+g^2\Phi^2\right)\chi_{\vec{k}} = 0,
\end{equation}
are then damped harmonic oscillators with a time-dependent mass.  In the case of a sinusoidally varying $\Phi$, we can reduce Eq.~\eqref{chimode}, after ignoring the expansion of the universe, to the Mathieu equation and predict the spectrum of amplifications.  If we allow $\Phi$ to be a sawtooth function whose amplitude decreases and whose period increases, the consequences for preheating are unclear.  On one hand, we expect that the time-varying period of oscillation should do harm to the period of parametric resonance. Some modes will experience small amplifications during each oscillation, but there is no assurance that any particular mode is amplified repeatedly.

\begin{figure}[b]
\includegraphics[width=\columnwidth]{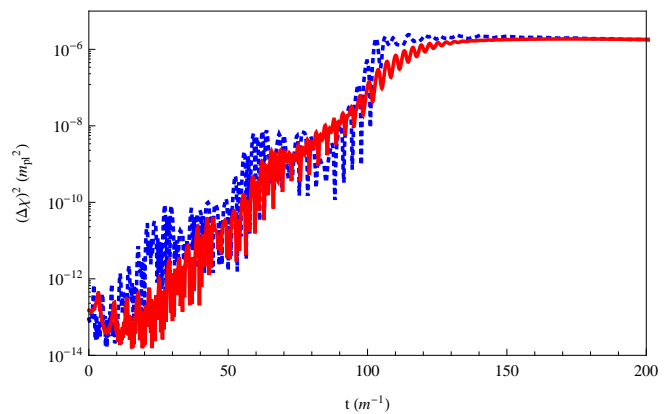}
\caption{Variance of the matter $\chi$ fields.  The red (solid) lines represent the fields when $\lambda/\mu^4=5000$ (initial $\gamma=8.7$), and the blue (dotted) lines represent the canonical ($f\rightarrow 0$) case.}
\label{fig:chicomparison}
\end{figure}

On the other hand, the sawtooth is actually many Fourier modes; it represents many forcing terms, each with different resonance bands.  Since the resonance in this case is much broader, we might expect that more modes are in resonance at any given time and the efficiency of parametric resonance is increased.  Figure~\ref{fig:chicomparison} shows the comparison between the canonical and the DBI cases.  The $\chi$ field is efficiently amplified during the preheating process, not hindered by the sawtooth oscillations. 

\begin{figure}[htb]
\includegraphics[width=\columnwidth]{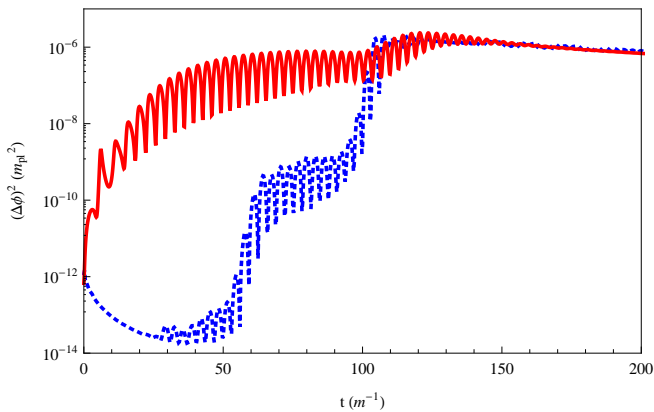}
\caption{Variance of the inflaton $\phi$ field.  The red (solid) lines represent the fields when $\lambda/\mu^4=5000$ (initial $\gamma=8.7$), whereas the blue (dotted) lines represent the canonical ($f\rightarrow 0$) case.}
\label{fig:phicomparison}
\end{figure}

Perhaps more interesting is the burgeoning importance of self-resonance.  The extra terms in Eq.~\eqref{DBIeom} give rise to self-interactions that provide a mechanism for self-resonance.  Unlike in the canonical case, the modes of $\phi$ undergo strong self-resonance in the presence of nonminimal terms.  Figure~\ref{fig:phicomparison} shows the dramatic difference in the two regimes; indeed, for the case of interest here, we see that the self-resonance is faster and more efficient than the induced parametric resonance in the matter field.  This can be seen in Fig.~\ref{fig:phicomparison}, since the variance of $\phi$ for the nonminimal case grows quickly in the early stages of the simulation, whereas in the canonical case, the variance of $\phi$ decreases during this time.  The existence of self-resonance is generic for different values of $\lambda/\mu^4$, becoming more significant as we depart further from the canonical case.

In both simulations, the periods of resonance cease around $t \sim 25\,m^{-1}$.  At this point, the two fields are sufficiently inhomogeneous that they begin to interact and thermalize via the $g^2\phi^2\chi^2/2$ interaction.  Lack of three-leg interactions, as is common in preheating simulations, makes it impossible for the $\phi$ field to decay completely \cite{Podolsky:2005bw}.  Nonetheless, we see a final state consistent with any simulations of preheating after quadratic inflation.

{\sl Discussion.--}
The above simulations, the first to model a scalar field obeying an equation of motion with nonminimal kinetic terms in a (3 + 1)-dimensional universe, reveal that even when those noncanonical terms are large, preheating occurs.  We see that the parametric amplification of nonthermal modes of a coupled matter field is just as efficient as that in the traditional preheating scenario.  Moreover, as we depart from the canonical case, parametric resonance is no longer the primary cause of preheating; instead, the self-interaction of the field causes almost immediate and efficient particle production.  The fact that the inflaton does not vary sinusoidally after inflation is not a death sentence for preheating; on the contrary, the spectral diversity of the homogeneously oscillating mode of $\phi$ acts as a source of inhomogeneity equally as efficient as a pure sinusoid.  

Since the structure of preheating is so similar, one can additionally expect that the gravitational wave spectrum from preheating \cite{Easther:2006gt,Easther:2006vd,Felder:2006cc,GarciaBellido:2007dg,Easther:2007vj,Dufaux:2007pt,GarciaBellido:2007af,Price:2008hq,Dufaux:2008dn} as well as any other observable consequences, e.g., non-Gaussianities \cite{Enqvist:2004ey,Jokinen:2005by,Barnaby:2006km,Chambers:2007se,Bond:2009xx,Barnaby:2010ke}, should not be any less significant, although we leave these analyses for other work.  We also delay a discussion of how self-resonance effects are important when considering the generation of pseudostable, nonlinear structures in the inflaton field \cite{Amin:2011hj,Amin:2013ika}.

Just as importantly, this work serves as a proof of concept that (3 + 1)-dimensional lattice simulations can simulate fields with nonminimal kinetic terms.  This should pave the way for simulating tensor fields with nonminimal kinetic terms sourced by scalar fields.

In a future paper, we will present the results of simulations with a range of values of $\gamma$ at the end of inflation.  We also delay predictions for the spectrum of amplifications that one expects from this process.  

{\sl Acknowledgments.--}  We want to thank Mustafa Amin for very useful discussions.  H.~L.~C. and J.~T.~G. are supported by National Science Foundation Grant No. PHY-1068080 and a Cottrell College Science Award from the Research Corporation for Science Advancement.

\end{document}